\documentclass[]{aa}
\usepackage{graphicx}
\begin{document}

\title{ 
On the formation and evolution of magnetic chemically peculiar stars
in the solar neighborhood}
\author{H.~P{\"o}hnl\inst{1}, E.~Paunzen\inst{1}, H.M.~Maitzen\inst{1}}

\mail{ernst.paunzen@univie.ac.at}

\institute{Institut f{\"u}r Astronomie der Universit{\"a}t Wien,
           T{\"u}rkenschanzstr. 17, A-1180 Wien, Austria}

\date{Received 20 April 2005 / Accepted 19 June 2005}
\titlerunning{On the formation and evolution of magnetic CP stars}{}

\abstract{In order to put strict observational constraints on the 
evolutionary status of the magnetic chemically peculiar
stars (CP2) of the upper main sequence, we have investigated a
well established sample of galactic field CP2 objects within a 
radius of 200\,pc from the Sun in the (X,Y)
plane. In total, 182 stars with accurate parallax measurements from
the Hipparcos satellite
were divided into Si, SiCr and SrCrEu subgroups based on classification
resolution data from the literature. Primarily, it was investigated if the CP2
phenomenon occurs at very early stages of the
stellar evolution, significantly before these stars reach 30\,\% of their life-time
on the main sequence. This result is especially important for theories
dealing with stellar dynamos, angular momentum loss during the pre- as well
as main sequence and stellar evolutionary codes for CP2 stars. 
For the calibration of the chosen sample, the well-developed framework of the 
Geneva 7-color and Str{\"o}mgren $uvby\beta$ photometric system was used.
We are able to show that the CP2 phenomenon occurs continuously at the 
zero age main sequence for masses between 1.5 and 4.5\,M$_{\sun}$. The 
magnetic field strengths do not vary significantly during the evolution
towards the terminal age main sequence. Only the effective temperature
and magnetic field strength 
seem to determine the kind of peculiarity for those stars. We found several 
effects during the evolution of CP2 stars at the main sequence, i.e. there are
two ``critical'' temperatures where severe changes take place. There is
a transition between Si, SiCr and SrCrEu stars at 10000\,K whereas a
significant decrease to almost zero of evolved SrCrEu objects with masses
below 2.25\,M$_{\sun}$ at 8000\,K occurs. These conclusions have to be 
incorporated into
models that simulate the stellar formation and evolution of stars between
1.5 and 4.5\,M$_{\sun}$ in the presence of strong magnetic fields.
}
\maketitle

\section{Introduction}

In P{\"o}hnl et al. (2003) we have shown that young
magnetic chemically peculiar stars (CP2 herafter, according to the
notation by Preston 1974) of the upper main sequence 
exist in young open clusters (ages between 10 and 140\,Myr) of our Milky Way. 
Our result received significant support by the detection (Bagnulo et al. 2003) 
of a 14.5\,kG magnetic field for 
HD~66318 in NGC~2516 which was also an object of the study
by P{\"o}hnl et al. (2003).

The strong stellar magnetic fields of these objects together with
the diffusion of chemical elements depending on the balance between
gravitational pull and uplift by the radiation field through absorption
in spectral lines manifests in strong overabundances 
for heavy elements such as silicon, chromium, strontium and europium
compared to the Sun.

The most important aspect for CP2 phenomenon is 
the origin of the global stellar magnetic fields
for those objects. Two theories have been developed in this respect (Moss 1989):
\begin{itemize}
\item The {\it dynamo} theory based on the existence of a contemporaneous
dynamo operating in the convective core of the magnetic stars. This means
that the magnetic fields should only appear at evolutionary stages between
the zero and terminal age main sequence (ZAMS and TAMS hereafter), preferably
closer to the TAMS.
\item The {\it fossil} theory has two variants: the magnetic field is either
the slowly decaying relic of the frozen-in interstellar magnetic field
or of the dynamo acting in the pre-main sequence phase. However, according
to this model, strong magnetic fields should be present when a star reaches
the ZAMS.
\end{itemize} 
Several divergent interpretations of the evolutionary status
and their implication for magnetic field theories 
of CP2 stars can be found in the literature. We have
summarized them in more detail in P{\"o}hnl et al. (2003).

Our main goal for this paper is to investigate the evolutionary
status for the CP2 group in the solar neighborhood and to answer the question raised by
Hubrig et al. (2000) who stated that 1.) The distribution of CP2 stars of masses below 
3\,M$_{\sun}$ in the Hertzsprung-Russell-Diagram (HRD hereafter)
differs from that of the ``normal'' 
stars in the same temperature range at a high level of significance: magnetic stars are 
concentrated toward the center of the main sequence band and 2.) 
Magnetic fields appear only in stars that have already completed at least 
approximately 30\% of their main sequence life-time. 

We have investigated three subgroups (Si, SiCr and SrCrEu) of CP2 stars which are
unambiguously classified and have accurate parallax measurements. This sample includes
182 objects within a radius of 200\,pc from the Sun in the (X,Y)
plane. The astrophysical parameters
of these objects were derived taking into account the peculiar characteristics and all
results have been carefully compared with already published values.

\begin{figure}
\begin{center}
\includegraphics[width=85mm]{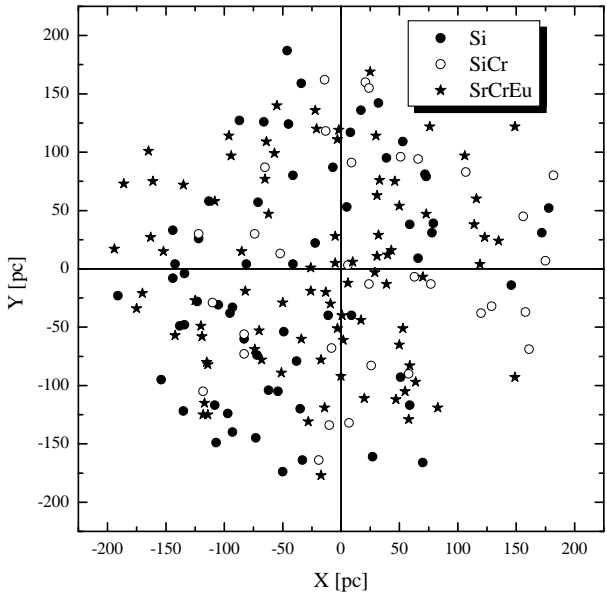}
\caption{The spatial distribution of the investigated sample of CP2 stars
divided into Si (filled circle), SiCr (open circle) and SrCrEu (asterisk) 
objects.}
\label{XY}
\end{center}
\end{figure}

\begin{figure*}[t]
\begin{center}
\includegraphics[width=175mm]{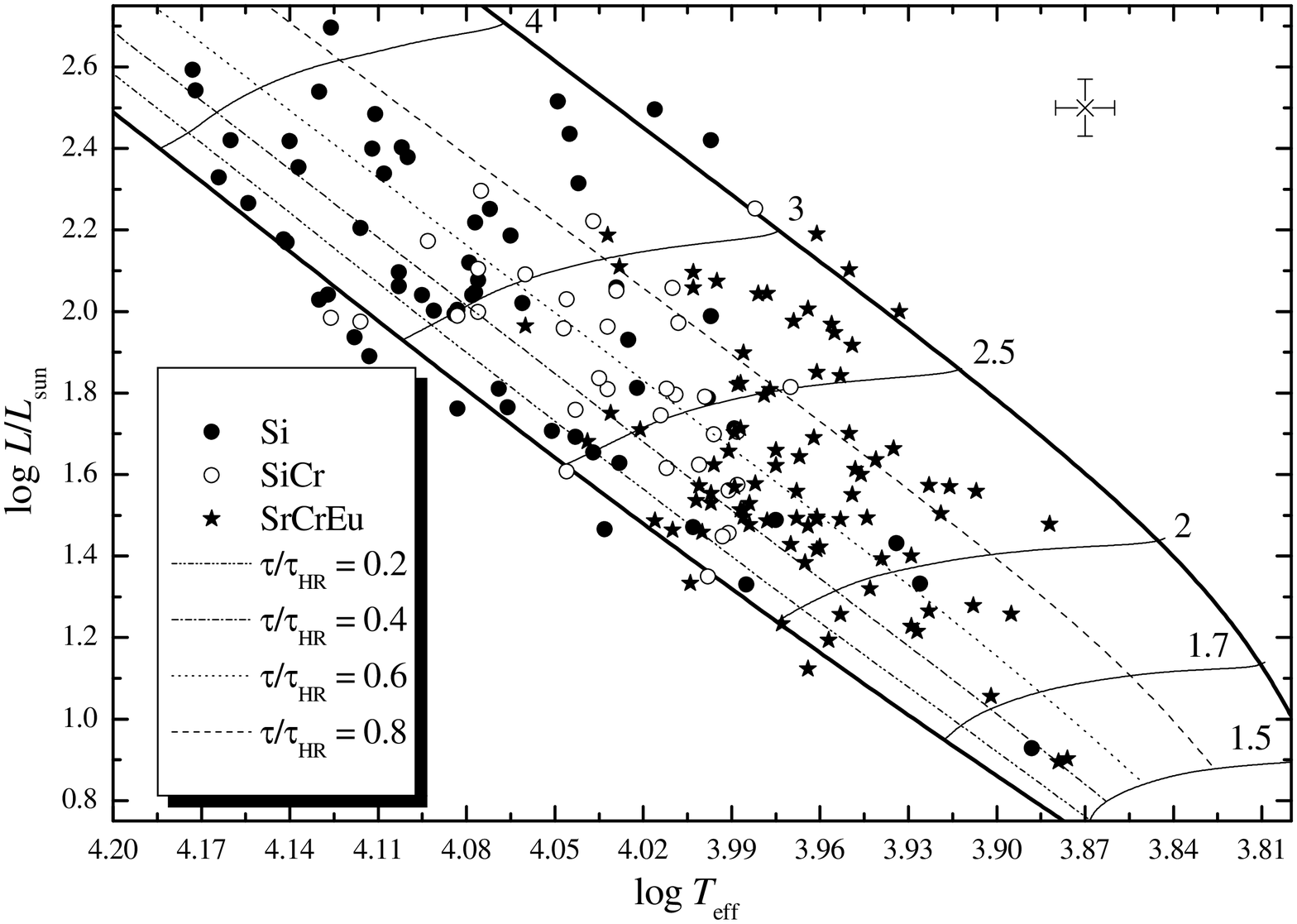}
\caption{The Hertzsprung-Russell-Diagram for our sample of CP2 stars,
the symbols are the same as in Fig. \ref{XY}. The evolutionary tracks from
1.5 to 4\,M$_{\sun}$, lines of
equal $\tau/\tau_{\rm HR}$, the ZAMS as well as TAMS
are for [Z]\,=\,0.016 taken from Schaller et al. (1992) and Schaerer et al.
(1993). The typical errors are plotted in the upper right corner.}
\label{hrd}
\end{center}
\end{figure*}

\section{Program stars}

The target selection for this analysis is based on two main criteria: 1.)
CP2 stars with known spectral characteristics and 2.)
Accurate parallax measurements available. We have taken the lists
of Renson (1991), G\'omez et al. (1998), Renson \& Catalano (2001),
Bychkov et al. (2003), Skiff (2003) and Paunzen et al. (2005) to search
for well-established CP2 stars. We have only included stars that are
unambiguously classified within these references. 

From this first list, objects were selected for which the absolute error
of the parallax is less than 17.5\% which is the upper limit for applying 
a correction as suggested by Lutz \& Kelker (1973). Our final list 
includes 182 objects which were divided into Si (61), SiCr (34) and
SrCrEu (87) stars. These are the three subgroups of the CP2 objects that are
described in more detail by Preston (1974) and Stepien (1994).
The phenomenology is due to classification resolution spectra in which
different lines of several elements are more pronounced than in normal
objects of the same effective temperature. Even if these subgroups
are investigated with high resolution spectra, the characteristic 
abundance pattern is still different (Wolff 1983). 

Figure \ref{XY} shows the spatial distribution of our sample. 
All objects are within a radius of 200\,pc from the Sun in the (X,Y)
plane. 

\section{Determination of the evolutionary status for the CP2 stars}

In the following subsections we describe the method to derive the 
log\,$L/L_{\sun}$ and log\,$T_{\rm eff}$ values for our program stars
in more detail. Special care was taken to include and correct the effects of the
spectral peculiarities when deriving the age as well as mass of the individual
stars. 

\subsection{Reddening and metallicity}

Our program stars are located within a radius of 200\,pc of the Sun in the (X,Y)
plane (Fig. \ref{XY}) for which
the interstellar reddening is not negligible any more. The most common way to derive the
reddening for single stars is to apply standard relations within the Str{\"o}mgren
$uvby\beta$ photometric system (Crawford 1975, 1979; Hilditch et al. 1983). 
These relations have to be used with caution when applied to CP2 stars because all 
calibrations are primarily based on the $\beta$ index. Due to variable 
strong magnetic fields, the $\beta$ index can give erratic values 
(Catalano \& Leone 1994). Another effect taken into account is
a ``blueing'' effect which manifests in bluer colors 
due to stronger UV absorption than in normal type stars which can be 
erratically interpreted as strong reddening (Adelman 1980).

We have therefore carefully checked the reddening
values from the Str{\"o}mgren $uvby\beta$ photometric system with other sources.

As a first approximation, we have used a ``standard''
value of 0.25\,mag\,kpc$^{-1}$ derived from open clusters (Dutra \& Bica
2000). These values were compared with the reddening laws listed by Chen et
al. (1998). As the authors note, the reddening for very close objects
can be overestimated. We found that for several objects the reddening derived
from Chen et al. (1998) has to be multiplied by a factor of 0.7 to be comparable
to the other sources. 

As a final step, we have compared and averaged the values from all sources.
As expected, many stars exhibit no reddening at all.

The metallicity has a severe influence on the determination of the age and
mass but only marginally affects the effective temperature (Glagolevskij 1994).
Classical CP2 stars are true Population I objects with intrisic solar 
abundances. Their peculiarities are restricted to the surface only
(Nishimura et al. 2004). It is therefore
justified to use solar isochrones for the determination of the age and mass
(G\'omez et al. 1998), but what metallicity should be chosen for the solar neighborhood?
Schaller et al. (1992) give [Z]\,=\,0.0188 whereas Allende Prieto et al. (2004) found that
the solar neighborhood would be ``metal-weak'' compared to 
the Sun by 0.06 to 0.11\,dex. 
We have therefore used a value of [Z]\,=\,0.016 for the isochrones which were
interpolated within the grids by Schaller et al. (1992, [Z]\,=\,0.02) and Schaerer et al.
(1993, [Z]\,=\,0.008).

\begin{figure*}[ht]
\begin{center}
\includegraphics[width=175mm]{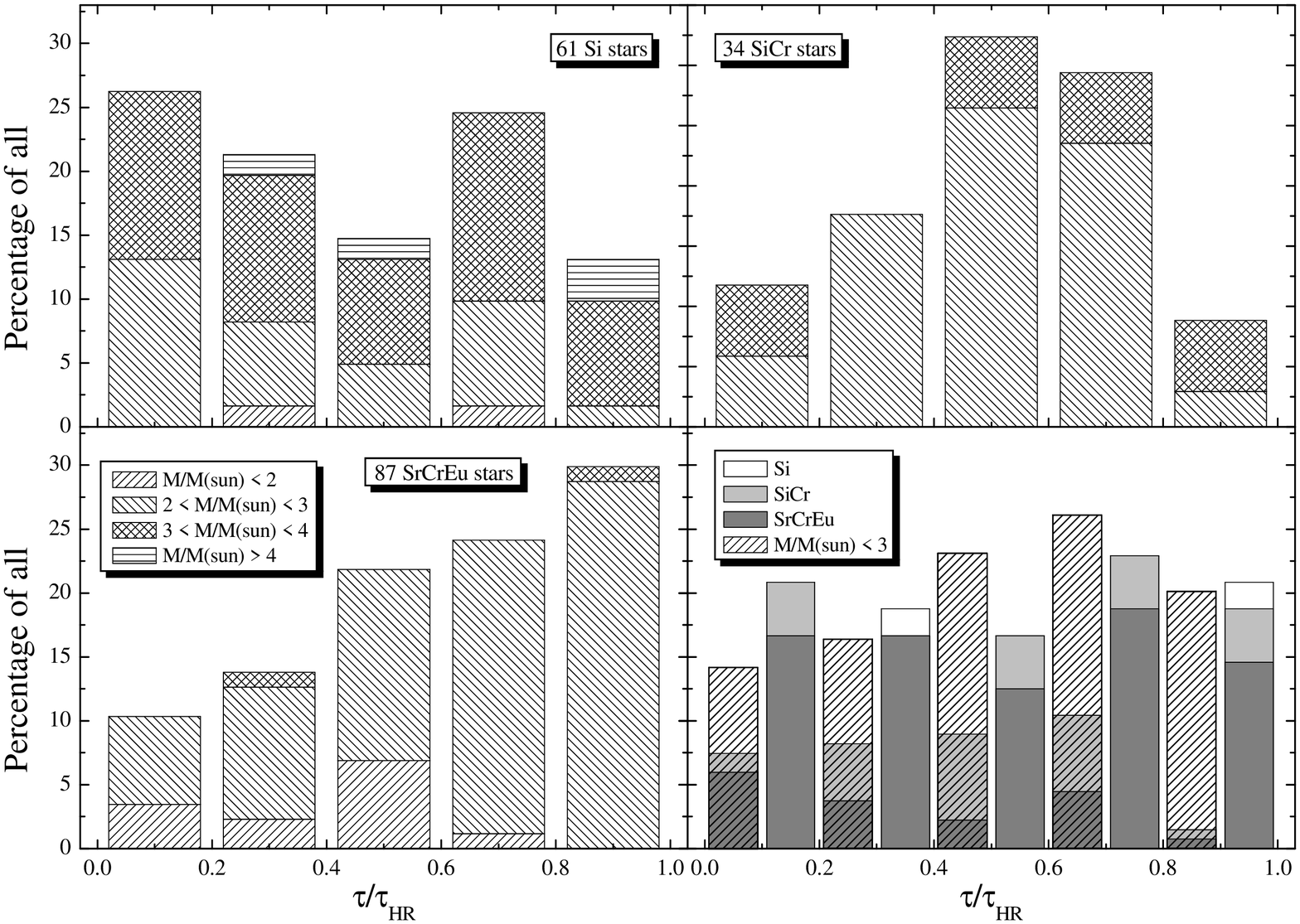}
\caption{A detailed statistical analysis of our sample. We
have divided the sample into Si (upper left), SiCr (upper right)
and SrCrEu (lower left) subgroups and investigated the incidence 
of these objects depending on their masses. The lower right panel
shows the comparison of all stars divided into subgroups with masses
higher (denoted with a pattern) and lower than 3\,M$_{\sun}$ according 
to the investigated sample by Hubrig et al. (2000). The Spearman rank
order correlation coefficient and the $\chi^2$ test
for these two samples give both a significance level of about 25\% for
a possible correlation. However,
these tests are only based on five data points.}
\label{hists}
\end{center}
\end{figure*}

\begin{figure*}
\begin{center}
\includegraphics[width=165mm]{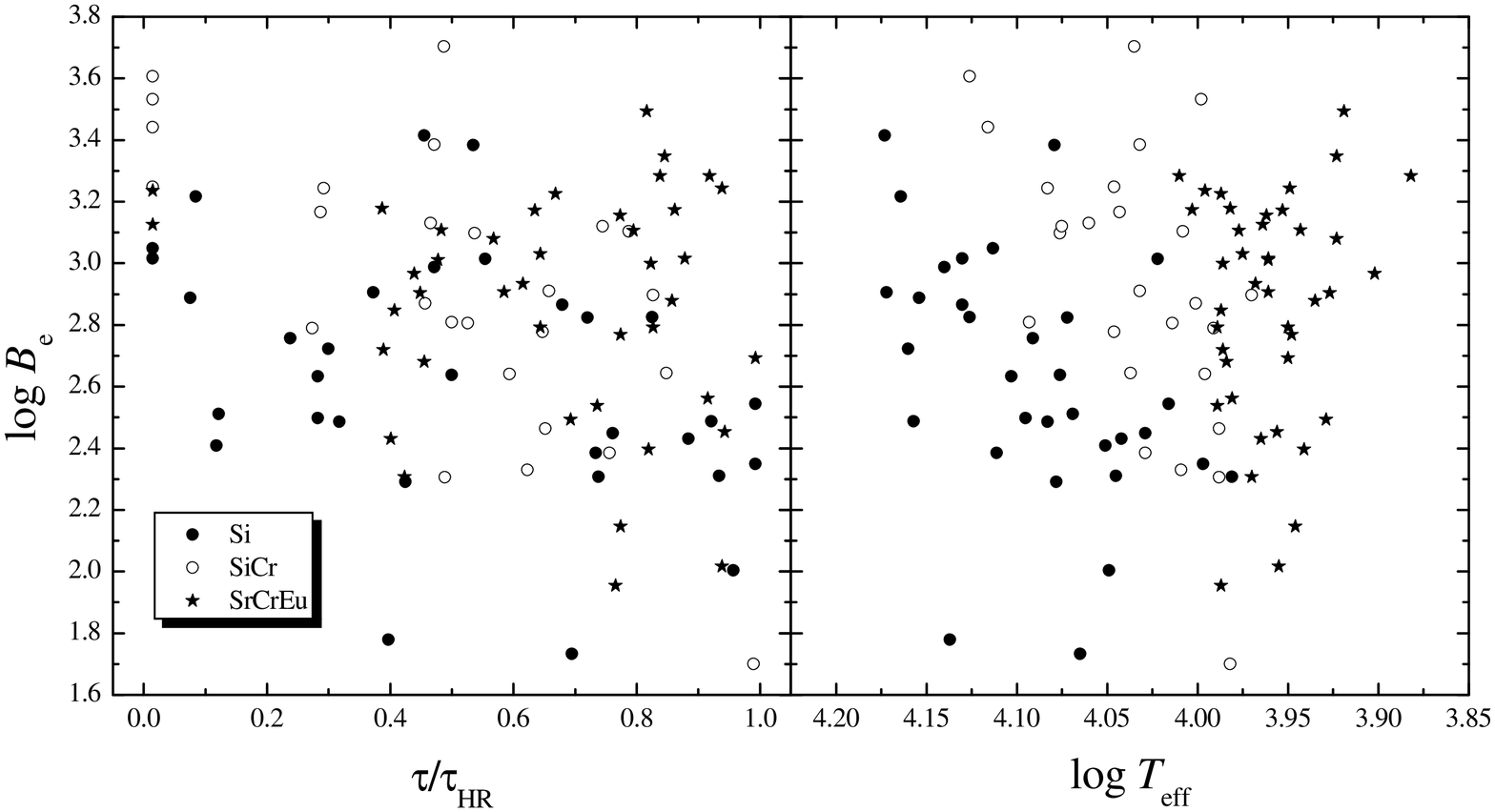}
\caption{The log\,$B_e$ versus $\tau/\tau_{\rm HR}$ 
as well as log\,$T_{eff}$ diagrams for our program stars.
The magnetic field strengths were taken from Bychkov et al. (2003).}
\label{be}
\end{center}
\end{figure*}

\subsection{Effective temperature and luminosity}

For the calibration of the effective temperature, we have made
use of the Geneva 7-color photometric system (Golay 1980).
All photometric measurements were taken from the General Catalogue of 
Photometric Data (GCPD, Mermilliod et al. 1997). 

The approach by Hauck \& K\"unzli (1996) was followed. For hot
CP2 stars ($T_{eff}$\,$>$\,9500\,K), the effective temperature
is given as
$$T_{eff} = -230 + 0.941\cdot T(X,Y)$$
where $T(X,Y)$ has to be calibrated within the grids by K\"unzli et al. (1997).
For cooler objects, we have used the calibration by Hauck \& North (1993)
which is based on $(B2-G)_0$. For the few cool objects without available Geneva 
colors, we have used the Str{\"o}mgren $uvby\beta$ photometric calibration
by Napiwotzki et al. (1993) which is based on the reddening free [$c_1$]
index
$$5400/T_{eff} = 0.2489 + 0.2698\cdot [c_1].$$
A comparison of the results for both 
calibrations of other CP2 stars in the same temperature range shows no
systematic trend which is in line with the results by Glagolevskij (2002).

The estimation of the errors for such a calibration has been widely 
discussed in the literature (Smalley \& Kupka 1997). Hauck \& K\"unzli (1996), for example,
list a statistical error for the photometric calibration of a large
sample of normal type stars of $\pm$386\,K depending on the temperature range.
North (1998) investigated especially CP2 stars and found an error
of the mean of 4.4\% and 3.4\% ($\Delta$\,log\,$T_{\rm eff}$\,=\,0.019 and 0.014)
for objects hotter and cooler than 9500\,K, respectively. We were not able to compare 
our calibrated effective temperatures with those used by G\'omez et al. (1998)
because the latter are not available.

The luminosity calibration for CP2 stars is also affected by stronger
UV absorption which results in correction factors $\delta_{BC}$ of up
to 0.3\,mag for the estimation
of bolometric magnitudes for this group (Lanz 1984, North 1998).
The luminosity can be calculated as:
$$\log L/L_{\sun} = 0.4\cdot (4.72 -M_V -BC + \delta_{BC})$$
with 
$$M_V = m_V + 5\cdot \log \pi +5 - 3.1\cdot E(B-V).$$
The values for the bolometric correction $BC$ and the standard solar value
were taken from Flower (1996). Finally, we have corrected for the
``Lutz-Kelker effect'' (Lutz \& Kelker 1973) which corrects
for the bias in the absolute magnitude of a star as estimated from
its trigonometric parallax. These corrections are rather small compared
to the error of the parallax measurements from Hipparcos at distances
up to 200\,pc from the Sun.

\subsection{Age and mass}

We have used interpolated isochrones with
[Z]\,=\,0.016 from Schaller et al. (1992) and Schaerer et al.
(1993) for the determination of the ages and masses for the program
stars.

The location of a star within the HRD is unambiguously defined by the
hydrogen concentration in the core $X_{\rm C}$ and the stellar mass.
The interpolation of these two parameters within the isochrones has been
described in P{\"o}hnl et al. (2003).

The knowledge of $X_{\rm C}$ allows us to determine the relative 
age $\tau/\tau_{\rm HR}$, where $\tau_{\rm HR}$ denotes the
time a star is on the main sequence. This is because
log\,$L/L_{\sun}$ and log\,$M/M_{\sun}$ are, at any stage of 
the stellar evolution, well correlated with $X_{\rm C}$
Schaller et al. (1992) list
value of $X_{\rm C}$\,=\,0.68 and 0.03 for the ZAMS 
($\tau/\tau_{\rm HR}$\,$\approx$\,0) and the
TAMS ($\tau/\tau_{\rm HR}$\,=\,1), respectively. These values are 
almost independent of the stellar mass. 

Our quadratic interpolation within the isochrones results in:  
$$\tau/\tau_{\rm HR} = 0.015 + 2.714\cdot(0.68 - X_{\rm C}) - 1.845\cdot(0.68 -
X_{\rm C})^2.$$
The statistical error for the complete sample, only taking
into account the errors within the interpolation grid, is less than $\pm$5\,\%
for the relative age. The errors for individual objects can be larger due
to the errors of the effective temperature (error of the statistical mean is less than
4.4\%, see P{\"o}hnl et al. 2003) and the luminosity (determined be the parallax error
which is always smaller than 17.5\%). But throughout this paper, only statistical
samples are used instead of values for individual CP2 stars which implies that 
the statistical error of less than $\pm$5\,\% for the relative age is justified.

\section{Results}

In Fig. \ref{hrd} the location of the program stars within the HRD 
is shown. We have included evolutionary tracks from 1.5 to 4\,M$_{\sun}$, 
lines of equal $\tau/\tau_{\rm HR}$, the ZAMS and TAMS
for [Z]\,=\,0.016 and the typical error bars. At first
sight it is evident that the CP2 stars occupy the whole width of the
main sequence which means that they are luminosity class V objects.

We performed a detailed statistical analysis
of the relative ages $\tau/\tau_{\rm HR}$ for the three samples
of Si, SiCr and SrCrEu objects. For this purpose we further divided
those samples according to the masses into four subsamples ranging 
from below 2\,M$_{\sun}$ to masses higher than 4\,M$_{\sun}$. 

Figure \ref{hists} shows the result graphically. The different groups
were binned with bin sizes of 0.2 for $\tau/\tau_{\rm HR}$. A smaller
bin size is not justified because of the overall errors and the additional 
introduction of small number statistics. From 
Figs. \ref{hrd} to \ref{be} we are able to conclude:
\begin{itemize}
\item The CP2 phenomenon occurs already at the ZAMS.
The percentage of all investigated CP2 stars with $\tau/\tau_{\rm HR}$\,$<$\,0.2 is
14.4\% as well as 20.8\% for objects below and above 3\,M$_{\sun}$ which accounts
for 16\% (or 29 stars) of the complete sample.
\item The magnetic field strengths do not vary significantly 
during the main sequence evolution.
\item The age distribution of objects below (no clear
trend with the age) and above (clear peak at 
0.6\,$<$\,$\tau/\tau_{\rm HR}$\,$<$\,0.8) 3\,M$_{\sun}$ is
statistically different. The Spearman rank
order correlation coefficient and the $\chi^2$ test (Rees 1987)
for these two samples both give a significance level of about 25\% for
a possible correlation. This result has to be taken with caution because
only five data points are include in this analysis.
\item The temperature and the magnetic field strength 
are the main parameters that determine the manifestation
of the chemically peculiarity, i.e silicon - chromium - strontium and
europium (from hotter to cooler temperatures)
\item Si stars: they can be found within the whole mass range, but more so
at young relative ages with higher masses. Their incidence seems to 
decrease with age except for
a peak at between 0.6\,$<$\,$\tau/\tau_{\rm HR}$\,$<$\,0.8. We are not able
to decide whether this is due to poor number statistics or a physical effect.
\item SiCr stars: most of these objects have masses between 3 and 4\,M$_{\sun}$.
They seem to be transition objects between Si and SrCrEu stars. There is,
for example, an
apparent edge at [log\,$T_{eff}$,log\,$B_e$] defined by [4.0,2.5] and [4.15,3.5] 
which clearly separates Si and SiCr stars. The age
distribution clearly peaks between 0.4\,$<$\,$\tau/\tau_{\rm HR}$\,$<$\,0.6.
\item SrCrEu stars: these are the low mass CP2 stars (M\,$<$\,3\,M$_{\sun}$)
with an upper limit of the effective temperature between 9500 and 10000\,K.
There is a clear negative correlation between the incidence and the age.
However, at least 24\% of all objects have relative ages below 0.4.
\end{itemize}
The question arises if the age distribution of CP2 stars with masses
below 3\,M$_{\sun}$ is significantly different from that of normal type objects in
the same spectral range as suggested by Hubrig et al. (2000). They have
used all apparent single type objects of the Bright Star Catalogue 
as a comparison which resulted in 412 objects. However, their CP sample
includes only 33 magnetic stars which is most certainly biased by
lower rotational velocities than for the normal type stars (Abt \& Morell
1995). The comparison
of such widely different samples has to be treated with caution. 
We have chosen to follow the approach by Paunzen et al. (2002) who used
the sample of the program stars to choose a random sample of normal type
objects by matching the effective temperature distribution. They
found for stellar masses below 2.7\,M$_{\sun}$ a peak of the relative
age distribution for 0.6\,$<$\,$\tau/\tau_{\rm HR}$\,$<$\,0.9 (see
Figure 2, Paunzen et al. 2002). We found exactly the same behaviour
for a random sample of normal type objects within a radius of 200\,pc 
of the Sun in the (X,Y) plane
which matches the sample of CP2 stars with masses below
3.0\,M$_{\sun}$. We therefore conclude that the different age distribution
of low mass CP2 stars as discussed by Hubrig et al. (2000) is due 
to a bias of the selection criteria and does not reflect an astrophysical
process.

North (1993) speculated whether a star can be born as a He-weak and
``die'' as a Si star, or, similarly, whether a Si star can become a Cr or a 
SrCrEu while evolving. From Fig. \ref{hists} it is obvious that there is a
surplus of SrCrEu objects near the TAMS. According to the models by
Schaller et al. (1992), an object with 2.5\,M$_{\sun}$, has at the ZAMS an
effective temperature of about 11150\,K, after 585\,Myr it reaches the TAMS
with an effective temperature of 8160\,K. During the life time on the ZAMS,
the stellar atmosphere changes due to rotational and magnetic effects
as well as due to the diffusion mechanism. These effects seem to manifest in the
HRD. There is a dramatic decline of evolved SrCrEu objects for masses lower than 
2.5\,M$_{\sun}$ whereas it is the opposite case for masses higher than that. There
are hardly any evolved Si stars but many evolved SrCrEu objects. We therefore
conclude that there are three different effects: 
\begin{enumerate}
\item Si stars change their peculiarity as they evolve towards temperatures cooler 
than 10000\,K on the main sequence
\item There is an obvious transition between Si, SiCr and SrCrEu objects at 10000\,K 
\item During the evolution
of CP2 stars with masses lower than 2.25\,M$_{\sun}$, there is a significant 
decrease of the incidence for peculiar objects at 8000\,K.
\end{enumerate}

\section{Conclusions}

We have investigated a sample of 182 CP2 stars for which homogeneous classifications
and accurate parallax measurements are available. The sample was further divided into
Si, SiCr and SrCrEu objects, the three main groups. For all program
stars effective temperatures, luminosities and thus ages as well as masses 
were calibrated with the help of standard evolutionary models. For this
purpose we used data and calibrations within the Geneva
7-color and Str{\"o}mgren $uvby\beta$ photometric system. Special care was
taken to include the appropriate corrections for the spectral and thus
flux peculiarities. 

The evolutionary status of CP2 stars is especially interesting 
because it still not clear whether these objects show their
peculiar nature soon after arriving at the ZAMS or
after about 30\,\% of their life time on the main sequence. Answering
this question from the observational side has a direct impact on the
theories about the origin of the strong stellar magnetic fields. 
One model describes the survival of frozen-in fossil fields originating from the
medium out of which the stars were formed, the other one 
follows the idea that a dynamo mechanism is acting in the interior of
these stars.

From the location of our program stars we are able to conclude that
the CP2 phenomenon occurs already at the ZAMS.
At least 16\% of the investigated CP2 stars have relative ages below
20\%. We find no significant variation of the magnetic field strengths  
over the whole main sequence. The effective temperature and the
magnetic field strength are the main 
parameters that determine the kind 
of the chemical peculiarity, i.e silicon - chromium - strontium and
europium (from hotter to cooler temperatures). 

The difference in the samples of CP2 stars with masses
above and below 3\,M$_{\sun}$ reported by Hubrig et al. (2000) seems 
to be a selection and thus bias effect of the peculiar and normal
sample. 

We find a strong decline of evolved SrCrEu objects for masses lower than 
2.5\,M$_{\sun}$, whereas there are few evolved Si stars but many 
evolved SrCrEu objects above this stellar mass. This can be interpreted
as Si stars changing their peculiarity 
as they evolve at temperatures cooler than 10000\,K which results in a
transition between Si, SiCr and SrCrEu objects at that point. 
Furthermore, the incidence of CP2 stars with masses lower than 
2.25\,M$_{\sun}$ decrease to almost zero below effective temperatures 
of 8000\,K.

These strict observational constraints have to be included and explained
by theories that model the stellar formation and evolution of stars between
1.5 and 4.5\,M$_{\sun}$ in the presence of strong magnetic fields.

\begin{acknowledgements}
This research was performed within the projects  
{\sl P17580} and {\sl P17920} of the Austrian Fonds zur F{\"o}rderung der 
wissen\-schaft\-lichen Forschung (FwF).
Use was made of the SIMBAD database, operated at the CDS, Strasbourg, France and
the NASA's Astrophysics Data System.
\end{acknowledgements}


\begin{thebibliography}{}
\bibitem[]{} Abt, H. A., \& Morrell, N. I. 1995, ApJS,99, 135
\bibitem[]{} Adelman, S. J. 1980, A\&A, 89, 149
\bibitem[]{} Allende Prieto, C., Barklem, P. S., Lambert, D. L., Cunha, K.
2004, A\&A, 420, 183
\bibitem[]{} Bagnulo, S., Landstreet, J. D., Lo Curto, G., Szeifert, T., Wade, G. A.
2003, A\&A, 403, 645
\bibitem[]{} Bychkov, V. D., Bychkova, L. V., Madej, J. 2003, A\&A, 407, 631
\bibitem[]{} Catalano, F. A., \& Leone, F. 1994, A\&AS, 108, 595
\bibitem[]{} Chen, B., Vergely, J. L., Valette, B., Carraro, G. 1998, A\&A, 336, 137
\bibitem[]{} Crawford, D. L. 1975, AJ, 80, 955
\bibitem[]{} Crawford, D. L. 1979, AJ, 84, 1858
\bibitem[]{} Dutra, C. M., \& Bica, E. 2000, A\&A, 359, 347
\bibitem[]{} Flower, P. J. 1996, ApJ, 469, 355
\bibitem[]{} Glagolevskij, Yu. V. 1994, Bull. Spec. Astrophys. Obs., 38, 152
\bibitem[]{} Glagolevskij, Yu. V. 2002, Bull. Spec. Astrophys. Obs., 53, 33
\bibitem[]{} Golay, M. 1980, Vistas in Astronomy, 24, 141
\bibitem[]{} G\'omez, A.E., Luri, X., Grenier, S., et al. 1998, A\&A, 336, 953
\bibitem[]{} Hauck, B., \& K\"unzli, M. 1996, Baltic Astronomy 5, 303
\bibitem[]{} Hauck, B., \& North, P. 1993, A\&A, 269, 403
\bibitem[]{} Hilditch, R. W., Hill, G., Barnes, J. V. 1983, MNRAS, 204, 241
\bibitem[]{} Hubrig, S., North, P., Mathys, G. 2000, A\&A, 539, 352
\bibitem[]{} K{\"u}nzli, M., North, P., Kurucz, R. L., Nicolet, B. 1997, A\&AS, 122, 51
\bibitem[]{} Lanz, T. 1984, A\&A, 139, 161
\bibitem[]{} Lutz, Th. E., \& Kelker, D. H. 1973, PASP, 85, 573
\bibitem[]{} Mermilliod, J.-C., Mermilliod, M., Hauck, B. 1997, A\&AS, 124, 349
\bibitem[]{} Moss, D. 1989, MNRAS, 236, 629
\bibitem[]{} Napiwotzki, R., Sch\"onberner, D., Wenske, V. 1993, A\&A, 268, 653
\bibitem[]{} Nishimura, M., Sadakane, K., Kato, K., Takeda, Y., Mathys, G.
2004, A\&A, 420, 673
\bibitem[]{} North, P. 1993, ASPC, 44, 577
\bibitem[]{} North, P. 1998, A\&A, 334, 181
\bibitem[]{} Paunzen, E., St{\"u}tz, Ch., Maitzen, H. M. 2005, A\&A, in press
\bibitem[]{} Paunzen, E., Iliev, I. Kh., Kamp, I., Barzova, I. S. 2002, MNRAS, 336,
1030
\bibitem[]{} P{\"o}hnl, H., Maitzen, H. M., Paunzen, E. 2003, A\&A, 402, 247
\bibitem[]{} Preston, G. W. 1974, ARA\&A, 12, 257 
\bibitem[]{} Rees D. G., 1987, Foundations of Statistics, Chapman \& Hall, London
\bibitem[]{} Renson, P. 1991, Catalogue G\'en\'eral des Etoiles Ap et Am,
Institut d'Astrophysique Universit\'e Li\`ege, Li\`ege
\bibitem[]{} Renson, P., \& Catalano, F. A. 2001, A\&A, 378, 113 
\bibitem[]{} Schaerer, D., Meynet, G., Maeder, A., Schaller, G. 1993, A\&AS, 98, 523
\bibitem[]{} Schaller, G., Schaerer, G., Meynet, G., Maeder, A. 1992, A\&AS, 96, 269
\bibitem[]{} Skiff, A. B. 2003, VizieR On-line Data Catalog: III/233. Originally published in: 
Lowell Observatory (2003)
\bibitem[]{} Smalley, B., \& Kupka, F. 1997, A\&A, 328, 349
\bibitem[]{} Stepien, K. 1994, In: Chemically Peculiar and Magnetic Stars, Zverko J., 
Ziznovsky J. (eds.), Slovak Academy of Sciences, p. 8
\bibitem[]{} Wolff, S. C. 1983, The A-type stars: Problems and perspectives, NASA SP-463
\end{thebibliography}
\end{document}